# Study on PM emissions in cellular automata model with slow-to-start effect

Qiao Yan-feng[a]   Xue Yu[a,b]   Wang Xue[a]   Cen Bing-ling[a]   Wang Yi[a]

[a]*Institute of Physical Science and Technology, Guangxi University, Nanning 53004, China*
[b]*Key Lab Relativist Astrophys, Nanning 530004, Guangxi, China*

**Abstract**

Based on the empirical particulate emission model, we studied Particulate Matter (PM) emission of some typical cellular automata VDR model and TT model with slow-to-start rules under periodic condition and open boundary condition. By simulations, it is found that the emission of the slow-to-start rule model reaches the maximum emission at metastable state under periodic boundary condition. Under open boundary condition, the phase diagram to reflect traffic congestion is obtained. The injection probability and removal probability have a great impact on PM emissions. Moreover, the effects of motion status on emissions in the VDR model and TT model are studied under two different boundary conditions. Numerical simulation shows that the PM emission of decelerating traffic flow reaches the maximum in the congestion state under periodic boundary condition. Under the open boundary conditions with the same departure probability, the variation trend of particulate emission varies with the state of different vehicle movements. For different removal probabilities, the different maximum concentration of particulate matter is emitted even if the vehicle is in the same motion state.

## 1. Introduction

In recent years, traffic congestion, traffic accidents and other traffic phenomena are becoming more and more serious with the increasing number of vehicles. In the metropolises, due to the motor vehicle fleet and the increasing emissions of toxic pollutants by industrial sources, air quality has become increasingly terrible. The high concentrations of harmful substances caused low visibility and various respiratory problems. The particulate matter (PM) has serious harm to human cardiovascular system and respiratory system[1,2].

Scientists have proposed various traffic flow models to study the real traffic flow evolution mechanism and formation of traffic congestion, such as continuity model, car-following model, cellular automata model and fluid dynamics model and so on[3-10]. Considering the influence of instantaneous velocity and instantaneous acceleration, Ahn proposed a hybrid regression model to predict vehicle emissions and fuel consumption[11]. Treiber and Kesting studied the instantaneous fuel consumption which includes vehicle properties, engine properties, and gear-selection schemes and implement it for different passenger car types representing the vehicle fleet. They concluded that the influence of congestions on fuel consumption is distinctly lower than that on travel time[12]. Madani and Moussa simulated the fuel consumption and engine pollutant emission using cellular automata traffic model[13]. Tian et al. and Wen et al used cellular automata NaSch model and FI model to study the energy dissipation of mixed traffic, respectively. The energy dissipation is affected by vehicle length, the maximum velocity and the mixing ratio of mixed traffic[14,15]. Tang et al., used the car-following model to analyze the impacts of fuel consumption and emissions on the trip cost without late arrival at the equilibrium state and the effect if signal light on fuel consumption and emissions[16,17] and discuss the relation of the trip cost with fuel consumption of vehicles in single-lane traffic and two-lane traffic [18-20]. Zhu et al. investigated reducing CO2 emission of traffic flow via delay-feedback control strategy[21]and carried out a numerical simulation for traffic additional emission on a signalized road[22]. More recently, Pan et al. simulated traffic PM emissions by using cellular automaton Nagel–Schreckenberg (NaSch) model with open boundary condition. Results indicate that in the free flow, low velocity limit was more energy conservative and environmentally friendly, while the high velocity limit was better when the traffic system became jammed[23]. Wang et al. also proposed the mixed cellular automaton NaSch model to study the particulate matter and other pollutants emission in mixed traffic flow in details. The numerical results illustrate that the emission of mixed traffic flow increases with the increase of vehicle mixture ratio, the maximum velocity and the number of long vehicles. Moreover, decelerating vehicles emit the most emissions[24].

In this paper, we study Particulate Matter(PM) emission of the typical cellular automata traffic models with slow-to-start rule(VDR model,BJH model and TT model) by combining the model of empirical particulate emission under two boundary conditions: periodic boundary and open boundary respectively. The typical slow-to-start models of traffic flow and traffic flow emission model are given in Section 2. The effects of initial conditions, traffic congestion, metastable state and different motion states on particulate emissions are studied in Section 3 through numerical simulation and theoretical analysis under periodic boundary conditions. Section 4 numerically simulates the VDR model under open boundary conditions, the traffic flow phase diagram and current flow diagram are obtained. In addition, the emission of particulate matter(PM) under open boundary conditions is studied. Finally, some conclusions are yielded.

## 2. Model description

### 2.1 Models with slow-to-start rule

The classical NaSch model only reveals free flow phase and congestion phase[25], while the measured data show that metastable phenomena exist in the actual traffic flow[26-28]. In view of the shortcomings of NaSch model, the slow-to-start rule of CA traffic model was introduced. The typical CA traffic models is the random slowing down model related to velocity (called VDR model)[29]. The slow-to-start rules can reproduce the metastable phenomena in traffic. We briefly review the definition of VDR model. In the VDR model, time and space are discrete. The road is divided into cells which can be either empty or occupied by a vehicle with a velocity $v = 0, 1, \cdots, v_{max}$. Vehicles are numbered from $1, 2, \cdots, N$. In evolution step of VDR model, the random delay probability p is related to velocity[29]: $p = p(v)$. If $v = 0$, the random slowing down probability is $p(v) = p_0$; if $v > 0$, this probability corresponds to $p(v) = p$. When $p_0 \gg p$ and $v_{max} > 1$, metastable state is reproduced in the fundamental diagram and phase separation phenomena occurs in traffic pattern. When $p = p_0$, VDR model becomes NaSch model. The evolution step of VDR model is as follows:

Step 0. Determine the velocity-dependent randomization parameter p(v):

$$p(v) = \begin{cases} p_0 : for\ v = 0 \\ p\ \ : for\ v > 0 \end{cases} \qquad (1)$$

Step 1. Deterministic acceleration:

$$v(i, t + \frac{1}{3}) \to \min(v(i,t) + 1, v_{max}) \qquad (2)$$

Step 2. Deceleration:

$$v(i, t + \frac{2}{3}) \to \min(v(i, t + \frac{1}{3}), gap_i(t)) \qquad (3)$$

Step 3. Randomization:

if v(i,t)=0, randomization with probability $p_0$

$$v(i, t + 1) \to \max(v(i, t + \frac{2}{3}) - 1, 0) \qquad (4)$$

if v(i,t)>0, randomization with probability p.

$$v(i, t + 1) \to \max(v(i, t + \frac{2}{3}) - 1, 0) \qquad (5)$$

Step 4. Motion:

$$x(i, t + 1) \to x(i, t) + v(i, t + 1) \qquad (6)$$

The velocity and position of the $i^{th}$ vehicle at time step $t$ in the model are depicted by $v(i,t)$ and $x(i,t)$, respectively. The headway between $i^{th}$ and $(i+1)^{th}$ at time step t is expressed as $gap_i(t) = x(i+1,t) - x(i,t) - L_{vehicle}$, in which $L_{vehicle}$ is a vehicle length.

### 2.2 Traffic emission model

Panis[30] et al. obtained the emission empirical formula of vehicle related to instantaneous velocity and acceleration by using nonlinear multiple regression technique and measured data. The emission model has been applicable to the emission of vehicle exhaust from urban traffic network and high way. The emission $E_n(t)$ of the nth vehicle in time t is defined as follows.

$$E_n(t) = \max[E_0, f_1 + f_2 v_n(t) + f_3 v_n(t)^2 + f_4 a_n(t) + f_5 a_n(t)^2 + f_6 v_n(t) a_n(t)] \tag{7}$$

where $v_n(t)$ and $a_n(t)$ represent the instantaneous velocity and acceleration of the nth vehicle at time t, respectively. $E_0$ is a lower limit of emission (g/s) specified for each vehicle and pollutant type parameters. The pollutant emission parameters of each vehicle from $f_1$ to $f_6$ are obtained by nonlinear regression analysis. The parameter values are given in table 1[30].

**Table 1**

Parameters for Eq.

| Pollutant | Vehicle type | $E_0$ | $f_1$ | $f_2$ | $f_3$ | $f_4$ | $f_5$ | $f_6$ |
|---|---|---|---|---|---|---|---|---|
| PM | Diesel car | 0 | 0.00e+00 | 3.13e-04 | -1.84e-05 | 0.00e+00 | 7.50e-04 | 3.78e-04 |

3. **Numerical simulation under periodic boundary condition**

In this section, the VDR model under different initial distribution states(homogeneous distribution, random distribution and jams distribution) with the periodic boundary conditions are simulation. The length of the lane L=$10^4$ cells corresponding to actual road length 37.5km. The unit length is selected as a vehicle length ($L_{vehicle}$=1). There are N vehicles on the road. Thus, the global density is expressed as $\rho = N/L$. The average velocity $\bar{v}$ of vehicle is calculated as follows:

$$\bar{v} = \frac{1}{T}\frac{1}{N}\sum_{t=t_0+1}^{t_0+T}\sum_{i=1}^{N} v(i,t) \tag{8}$$

where $t_0$ is the relaxation time and T is the total evolution time. The simulation is performed with 50 independent runs in different initial configurations. Each runs for $3 \times 10^5$ iterations and the first $2 \times 10^5$ iterations are discarded in measuring quantities of interest.

**3.1 Fundamental diagram**

The fundamental diagram of VDR model from different initial conditions is shown in Fig.1(a). Fig.1 (a) shows the fundamental diagram from initial random distribution is same as those from initial compact blocking distribution. Fig.1 (b) shows the fundamental diagram of VDR model in homogeneous and stochastic initial distributions is compared with one of NaSch model. Obviously, there exists a metastable state in the fundamental diagram of VDR model.

The distribution of the three different initial distribution states is described in the following. For the initial condition of homogeneous distribution, vehicles are uniformly distributed on the road when vehicle density is less than 0.5( $\rho$ <0.5). When the density exceeds 0.5( $\rho$ >0.5), vehicles are first uniformly distributed according to the density of 0.5. Then, vehicles exceeding density 0.5 are randomly distributed among empty cells on the road(hom). The 2$^{nd}$ initial condition is distribution at random and vehicles are randomly distributed in an empty cell on the road(ran). The 3$^{rd}$ initial condition is a compact jamming distribution(jam). Initially, vehicles are allocated bumper-to-bumper to form a compact cluster from the first cell at the leftmost end on the road and the gap between the two adjacent vehicles is zero.

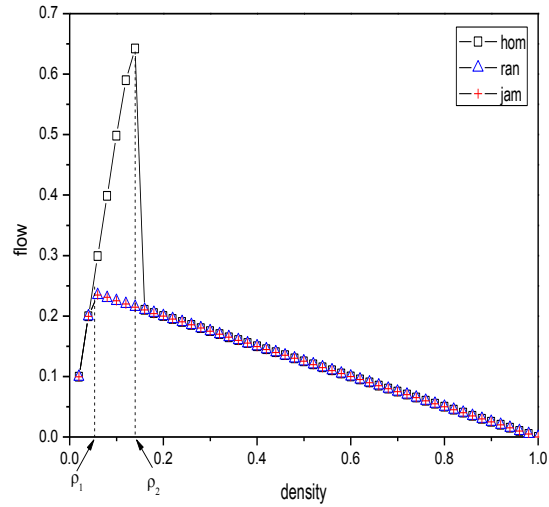

(a)

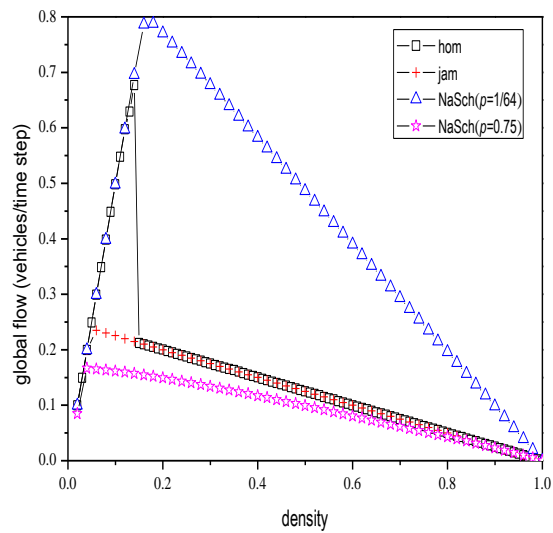

(b)

Fig.1. (a) Fundamental diagram of VDR model in different initial states, (b) Fundamental diagram of VDR model and NaSch mode($v_{max}=5, p=1/64, p_0=0.75, L=10000$).

**3.2 The average velocity**

Fig.2. is the average velocity of VDR model corresponding to Fig.1 (a). It can be found that the average velocity shows the different properties for homogenous and jamming initial conditions. The average velocity of jamming initial conditions gradually decrease at vehicle density $\rho$ =0.06. It implies that the traffic congestion occurs. However, for homogeneous initial distribution, the average velocity suddenly breaks down at vehicle density $\rho$ =0.14. It indicates the traffic phase transition occurs.

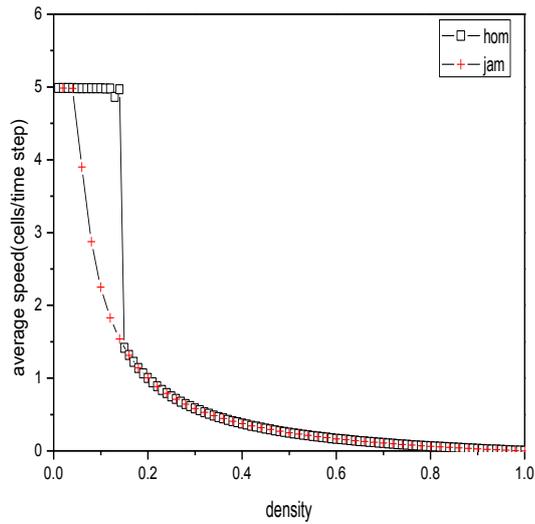

Fig.2. the average velocity of VDR model.

### 3.3 The spatio-temporal pattern

Fig.3. is the spatio-temporal pattern of VDR model, in which exhibits the evolution of vehicles and phase separation.

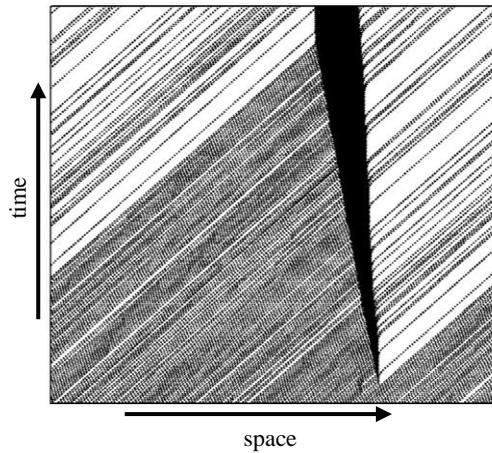

Fig.3. spatio-temporal of VDR model. (a) $\rho = 0.14, (v_{max} = 5, p = 1/64, p = 0.75)$

When the density is at $\rho_2 = 0.14$ (metastable state see fig.1 (a)), the flow of the system reaches its maximum and no jams occurs. However, studies have found that the metastable state of the VDR model is very sensitive to disturbances[31], and jams can be occur due to internal fluctuations and external disturbances, such as the stopping of a vehicle. Fig. 3 shows that at the metastable state ($\rho = 0.14$), the system will be blocked if one of the vehicles in the system stops moving at a certain time due to artificial disturbance.

### 3.4 Numerical simulation for emissions

The emissions of traffic flow can be reflected by the average emissions of vehicles. The computation of the average emissions is analogous to one of the average velocity. The average emission $\bar{E}$ is defined as follows:

$$\overline{E} = \frac{1}{T}\frac{1}{N}\sum_{t=t_0+1}^{t_0+T}\sum_{i=1}^{N} E_i(t) \tag{9}$$

Where the pollutant emission of the i$^{th}$ vehicle at time t is expressed as $E_i(t)$. The instantaneous emission of each vehicle is calculated according to Eq.(7), and then the average emission of traffic flow is calculated statistically by using Eq. (9).

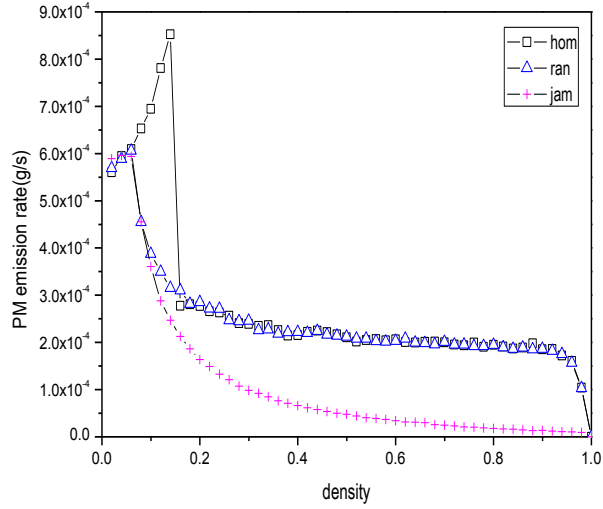

Fig.4. PM emission rate of VDR model in the different initial conditions.

Fig.4 shows the average PM emissions obtained by the VDR model under the different initial conditions. It can be found that the emissions of the VDR model are different under different initial conditions. The emissions gradually increase in free flow except for initial conditions of a compact jams distribution (jam). Under the initial condition of homogeneous distribution(hom), Fig.4 reveals the emissions sharply change at the density ρ =0.14, which corresponds to the phase transition from free flow in the metastable state to traffic congestion in the fundamental diagram of Fig.1(a). In the metastable region with densities ranging from ρ =0.06 to ρ =0.14，the emissions of traffic flow are always increasing until it suddenly drops at the phase transition point ρ =0.14. The PM emissions almost unchanged in density between ρ =0.2 to 0.9, then decreased after 0.9. For initial conditions of random distribution(ran), it can be seen from Fig.4 the emissions gradually increase and reach the maximum when traffic congestion occurs. For initial condition of a dense jams distribution (jam). In free flow traffic, emissions remain the same, while emissions in congestion decline and decrease.

### 3.5 Vehicle motion state effect on emissions

In this section, we study the emissions of VDR model in different states of motion, which are divided into acceleration($v_i(t+1)>v_i(t)$), deceleration($v_i(t+1)<v_i(t)$) and uniformly moving ($v_i(t+1)=v_i(t)$). Figs.5 (a),(b) and (c) show the average emissions of the acceleration mode, deceleration mode and uniformly moving, respectively. Fig.5 (a) shows the emission remains unchanged in the free flow for all the initial conditions. In the traffic congestion, the emissions reduces to about 0.030(g/s) at the density from $\rho$ =0.16 to 0.95 for initial condition of homogenous distribution(hom) and random distribution(ran) except for jams distribution(jams). Under the initial condition of jams distribution(jam), the emission keeps unchanged from free flow to congestion at density $\rho$ =0.8 and suddenly drops down after density $\rho$ =0.8. Fig.5 (b) is the emissions of deceleration mode for all the initial conditions. It is very obvious that the emissions from initial condition of homogenous distribution(hom) and random distribution(ran) increase with density and reach the maximum value 0.08(g/s). The emission of an initial dense jams distribution(jam) shows the minimum value. Fig.5 (c) exhibits the emission curves of uniform motion mode for all the initial conditions. It is clear that emissions trend of uniform motion mode is consistent with those of VDR model in Fig.4, but its order of magnitude is smaller than total emissions of VDR model in all the initial conditions. By comparing the emissions in Figs.5 (a),(b) and (c), it can be concluded that the emission is the largest in the congestion deceleration state.

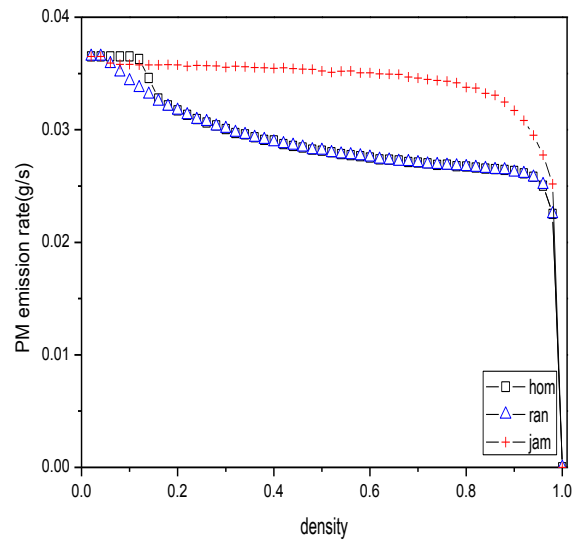

(a)

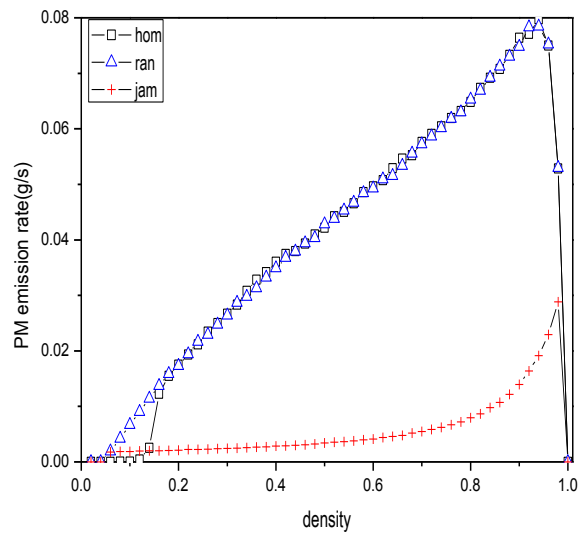

(b)

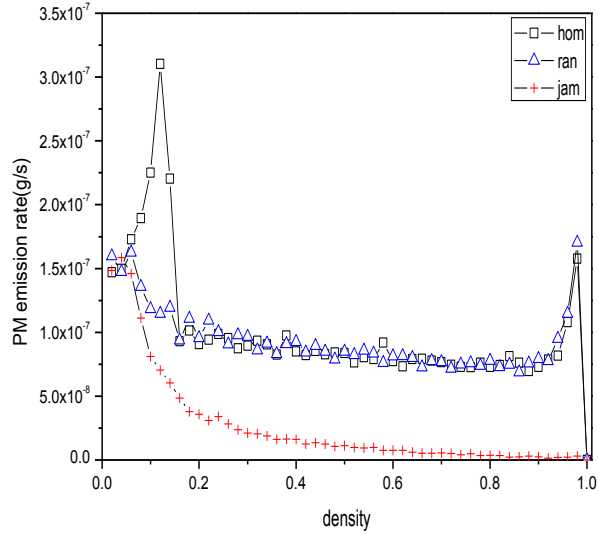

(c)

Fig.5. PM emission rate in accelerated mode(a), decelerated mode(b) and uniform motion(c) under periodic boundary conditions.

## 4. Numerical simulation under open boundary condition

So far, the traffic flow CA model with open boundary conditions has been widely studied[32-33]. In our model, the open boundary conditions are set as follows: out of the system ($i$=0) a vehicle with velocity $v_{max}$ is created with probability $α$. If $i$ =1 is occupied by another vehicle, the injected vehicle is removed. On the right boundary, the leading car is removed from the system with the probability $β$ when $x_{lead} > L$. Under the condition of open boundary, the length of road L=$10^4$ cells corresponding to actual road length 37.5km. The parameters of $p_0$ is set to 0.5, p is set to 0.1 and $v_{max} = 5$. The simulation is performed with 500 independent runs in different initial configurations. Each runs for $3 \times 10^5$ iterations and the first $2 \times 10^5$ iterations are discarded in measuring quantities of interest.

### 4.1 Phase diagram and flow diagram

The phase diagram of the VDR model is shown in Fig.6(a). The phase diagram of VDR model includes free flow phase, jamming phase and maximum current phase. Free flow and jamming phase are separated by a curve. After the injection probability $α > 0.45$ and the exit probability $β > 0.85$, the system will reach the maximum current phase.

In order to analyze the influence of left and right boundaries on the current under the open boundary condition, Fig.6(b) gives the variation diagram of the current with injection probability α under different exit probabilities $β$ ($β$=0.2, 0.5, 0.8, 1.0). The current diagram in Fig.6 (b) is in good agreement with the phase diagram in Fig.6 (a). For instance, when exit probability $β$=0.2, the current of the system increases with the increase of the injection probability $α$ in Fig.6 (b). However, after the injection probability is greater than 1.5, the current reaches the maximum and then remains unchanged, which indicating that the system reaches maximum capacity and there is congestion on the road. In Fig.6 (a), it is obvious that when exit probability $β$=0.2, injection probability $α$=0.15 is the cut-off point of free flow phase and jamming phase.

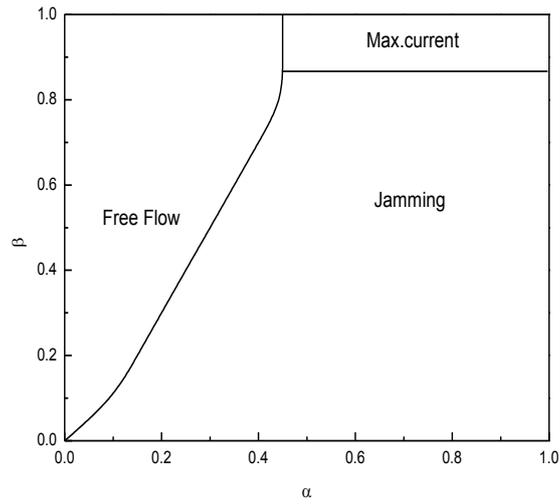

(a)

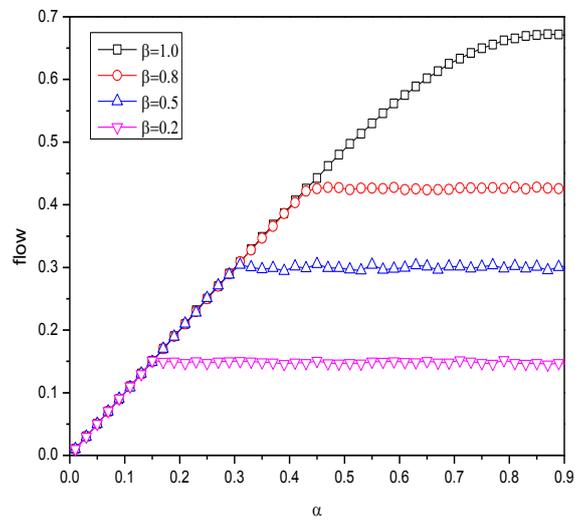

(b)

Fig.6. (a) The phase diagram of the VDR model, (b) The flow diagram

**4.2 Numerical simulation and analysis for emissions**

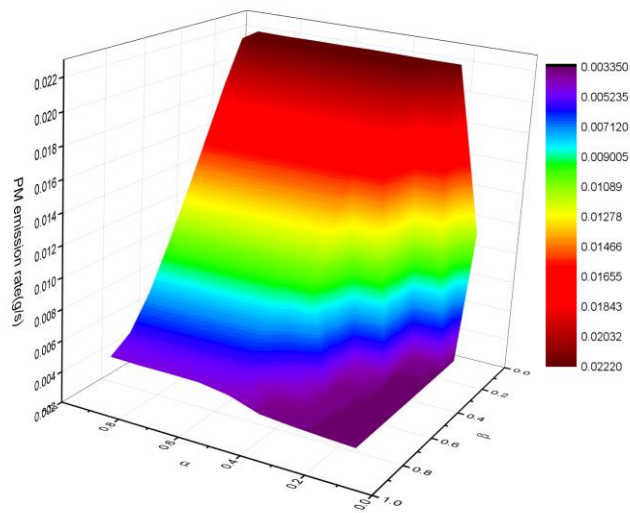

Fig.7. Profile of PM emissions under open boundary condition.

Fig.7 shows the three-dimensional diagram of average PM emissions of vehicles under different injection and exit probabilities. The bar on the right indicates that from top to bottom, emissions increase as the color deepens. It can be seen from the three-dimensional figure that injection probability and exit probability have great influence on emissions. When the injection probability is small and the exit probability is large, the emission of particulate matter is small, because there are fewer vehicles accelerating and decelerating in the system. On the contrary, when the probability of entrance is large and the probability of removal is small, the number of vehicles in the system will increase. When the maximum capacity of the system is exceeded, vehicles will become congested on the road. In order to prevent collisions, drivers need to constantly adjust the speed of vehicles, which makes vehicle emissions increase.

**4.3 Effect of vehicle motion state on emissions**

Similar to the periodic boundary conditions, the PM emissions in different motion states of VDR model under open boundary conditions is studied. Figs.8 (a),(b) and (c) show the average emissions of the acceleration mode, deceleration mode and uniformly moving, respectively. It can been see from Figs.8 that with the same exit probability $β$, the variation trend of vehicles' emissions in each motion state is different. Under the same motion state, for different exit probabilities $β$, the variation trend of emissions is also different.

Detailed analysis is now carried out for PM emissions in each motion state when the exit probability is 0.2($β$=0.2). In the free flow region($α$<0.15), the emissions of accelerating vehicles increase to maximum levels and remain unchanged, while that the emissions of decelerating vehicles and uniform vehicles is almost zero. Since there is no wide congestion in the system in the free-flow, there is enough safe distance between vehicles to allow the vehicles to travel at maximum speed on the road, thus the emissions of accelerating vehicles are maximal in the free-flow zone. When the injection probability is greater than 0.15($α$>0.15), the system changes from free flow phase to jamming phase (see Fig.6(a)), and congestion occurs on the road. Drivers reduce the velocity of vehicles to ensure safe driving, so there are more deceleration vehicles and fewer acceleration vehicles. Which makes the emissions of the decelerating vehicles increases and the accelerating vehicles decreases. However, it can be seen from Fig.6(b) that after the injection probability is greater than 0.15, the current of the system reaches its maximum and remains unchanged, which means that the proportion of accelerating and decelerating vehicles in the system unchanged, so the emissions of accelerating and decelerating vehicles also remain unchanged. In addition, when $α$>0.15, some vehicles in the system will travel at a constant speed with a lower velocity, so the emission of vehicles at a constant speed will increase after the occurrence of congestion. The emissions of vehicles with exit probability $β$=0.5 are similar to $β$=0.2. Nevertheless, when the case β=1, which means vehicles can simply move out of the system from the right boundary. There is no jams appear in system. According to the phase diagram in Fig.6(a), when β=1, the system only has free flow phase and maximum flow phase. Therefore, there are very few deceleration vehicles in the system, and the vehicle can travel through acceleration within maximum speed. Therefore, the emissions of accelerating vehicles are the most and decelerating vehicles are a little. When $β$=1, although there are many vehicles with uniform speed on the road, according to equation(7), the emission of vehicles with uniform speed is very small(see the right of Fig.8(c)).

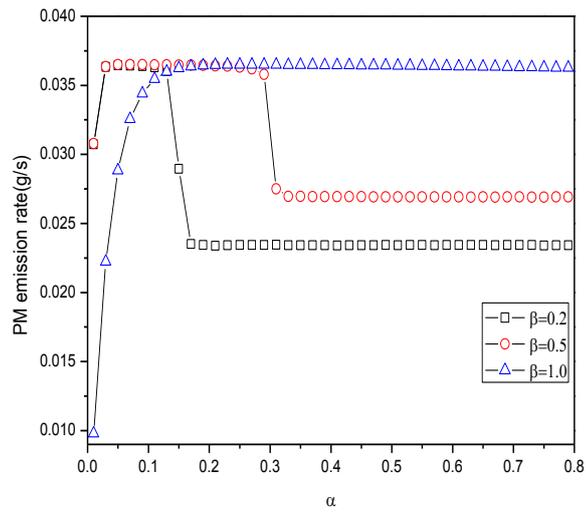

(a)

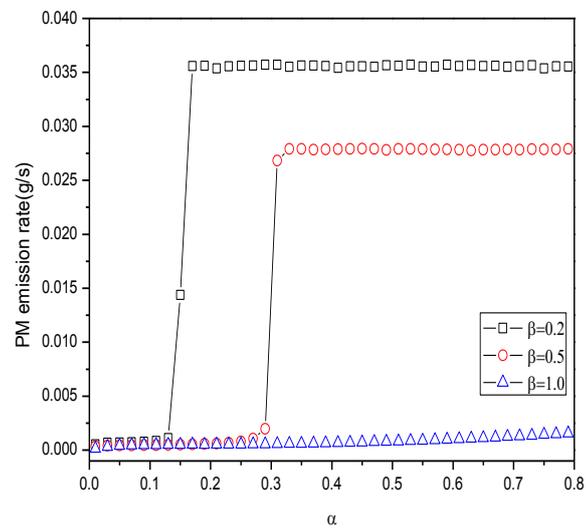

(b)

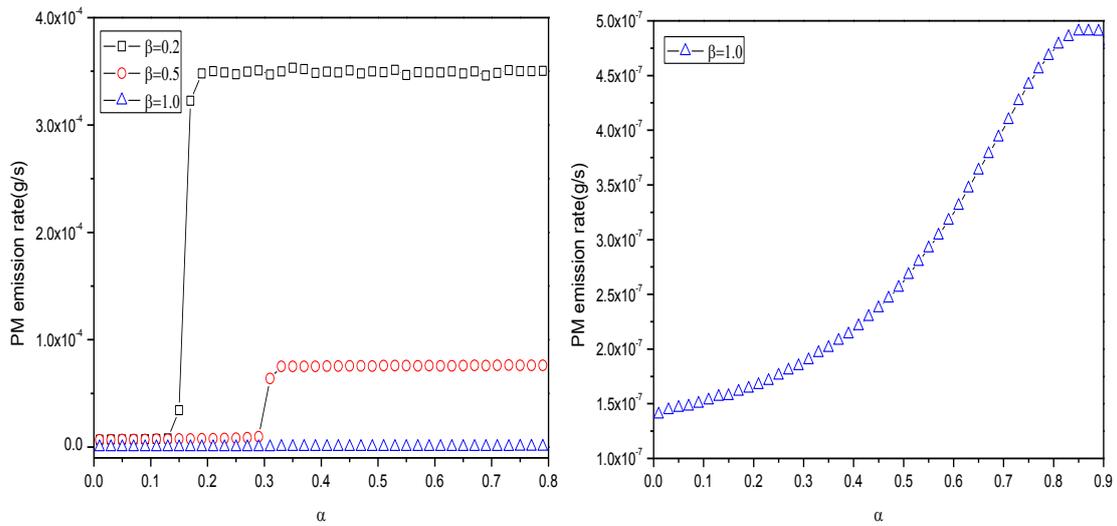

(c)

Fig.8. PM emission rate in accelerated mode(a), decelerated mode(b) and uniform motion(c)

under open boundary conditions.

## 5. Conclusions

In this paper, based on the empirical particulate emission model, we studied Particulate Matter (PM) emission of a typical cellular automata traffic model (VDR model) with slow-to-start rules under periodic boundary condition and open boundary condition.

The fundamental diagram of this model shows free-flow traffic, metastable state and traffic congestion under periodic boundary condition. Simulations illustrate that the initial conditions have a great impact on traffic emissions under periodic boundary condition. PM emission of VDR model from free-flow to metastable state rises to increase and reaches the maximum value at transition point $\rho \approx 0.14$ under the uniform initial condition. When traffic jamming occurs, PM emission of VDR model suddenly drops and then gradually decreases until traffic is completely congested and emissions approach zero. For the randomly distributed initial condition, the trend of traffic discharge is relatively monotonous, increasing gradually and then decreasing. For the initial conditions of dense congestion, traffic discharge remains constant during free flow and decreases when congestion occurs. Under open boundary condition, the phase diagram and flow diagram of VDR model are obtained. Numerical simulations reveal that the injection probability and removal probability affect PM emissions. Furthermore, the effects of motion states of vehicles on emissions in the VDR model are discussed respectively under periodic boundary condition and open boundary condition. The numerical simulation displays that PM emission of decelerated vehicles in congestion state achieves the maximum value and exceeds those of accelerated vehicles under periodic boundary condition. Therefore, for periodic boundary system, the initial conditions play a significant role in the formation of traffic congestion of traffic flow and PM emission of traffic. For open boundary system, the exit probability plays a major role in traffic congestion and particulate emissions. The small exit probability will result in traffic jamming. Under the same exit probability, the variation trend of particulate emission varies with the state of different vehicle movements. For different exit probabilities, the different maximum concentration of particulate matter is emitted even if the vehicle is in the same motion state.

## Acknowledgments

The project supported by the National Natural Science Foundation of China (Grant No. 11262003), the Natural Science Foundation of Guangxi, China (Grant No. 2018GXNSFAA138205), Guangxi Higher Education Undergraduate Teaching Reform Project （2019JGZ102）.

References

[1] M.Stafoggia, A.Faustini, M.Rognoni, et al., Air pollution and mortality in ten Italian cities. Results of the EpiAir Project. Epidemiologia e prevenzione 33 (2009) 65-76.

[2] C.Arden Pope III, Richard T.Burnett, Michael J.Thun, Eugenia E.Calle, Daniel Krewski, Kazuhiko Ito, George D.Thurston, Lung cancer, cardiopulmonary mortality, and long-term exposure to fine particulate air pollution, JAMA 287 (2002) 1132–1141.

[3] T.Nagatani, Modified KdV equation for jamming transition in the continuum models of traffic, Physica A 261 (1998) 599-607.

[4] G.F.Newell, Nonlinear Effects in the Dynamics of Car Following, Operations Research 9 (1961) 209-229.

[5] D.Helbing, Traffic and related self-driven many-particle systems, Rev. Modern Phys 73 (2001) 1067–1141.

[6] B.S.Kerner, H.Rehborn, Experimental Properties of Phase Transitions in Traffic Flow, Physical Review Letters 79 (1997) 4030-4033.

[7] K.Nagel, M.Schreckenberg, A cellular automaton model for freeway traffic, Journal de Physique I 2 (1992) 2221-2229.

[8] D.Chowdhury and L.Santen, A.Schadschneider, Statistical physics of vehicular traffic and some related systems,Phys.Rep 329 (2000) 199-329

[9] Nagatani, Takashi, The physics of traffic jams, Reports on Progress in Physics 65 (2002) 1331-1386.

[10] Kerner, S.Boris, Empirical macroscopic features of spatial-temporal traffic patterns at highway bottlenecks, Physical Review E 65 (2002) 046138.

[11] K.Ahn, H.Rakha, A.Trani, et al., Estimating Vehicle Fuel Consumption and Emissions based on Instantaneous Speed and Acceleration Levels, Journal of Transportation Engineering 128 (2002) 182-190.

[12] Martin Treiber, Arne Kesting, How much does traffic congestion increase fuel consumption and emission? Applying a fuel consumption model to the NGSIM trajectory data,Transp. Res. Board Meet. (2008) 1-17.

[13] Abdellah Madani, Najem Moussa, Simulation of fuel consumption and engine pollutant in cellular automaton, J. Theor. Appl. Inf. Technol. 35 (2) (2012) 250-257.

[14] Tian Huan-Huan, Xue Yu, Kan San-Jun, Liang Yu-Juan, Study on the energy consumption using the cellular automaton mixed traffic model, Acta Phys. Sin. 58 (2009) 4506-4513.

[15] Wen Jian, Tian Huan-huan, Kan San-Jun, Xue Yu, Study on the energy consumption of cellular automaton FI model for mixed traffic model, Acta Phys. Sin. 59 (11) (2010) 7693-7700.

[16] Tie-Qiao Tang, Tao Wang, Liang Chen, Hua-Yan Shang, Impacts of energy consumption and emission on the trip cost without late arrival at the equilibrium state, Physica A 479 (2017) 341-349.

[17] Tie-Qiao Tang, Zhi-Yan Yi, Qing-Feng Lin,Effects of signal light on the fuel consumption and emissions under car-following model, Physica A469 (2017) 200-205.

[18] Tie-Qiao Tang, Peng Liao, Hui Ou, Jian Zhang, A fuel-optimal driving strategy for a single vehicle with CVT, Physica A 505 (2018) 114-123.

[19] Tie-Qiao Tang, Tao Wang, Liang Chen, Hai-Jun Huang, Analysis of the equilibrium trip cost accounting for the fuel cost in a single-lane traffic system without lata arrival, Physica A 490

(2018) 451-457.

[20] Hui Ou, Tao Wang, Tie-Qiao Tang, Analysis of trip cost in a two-lane traffic corridor with one entry anf one exit, Physica A 524 (2019) 65-72.

[21] Zhang Li-dong, Zhu Wen-Xing, Delay-feedback control strategy for reducing CO2 emission of traffic flow system, Physica A 428 (2015) 481-492.

[22] Zhu Wen-Xing, Zhang Jing-Yu, An original traffic additional emission model and numerical simulation o a signalized road, Physica A 467 (2017) 107-119.

[23] Wei Pan, Yu Xue, Hong-Di He, Wei-Zhen Lu, Impacts of traffic congestion o fuel rate, dissipation and particle emission in a single lane based on Nasch Model, Physica A 503 (2018) 154-162.

[24] Wang Xue, Xue Yu, Cen Bing-ling, Zhang Peng, He Hong-di, Study on pollutant emissions of mixed traffic flow in cellular automaton, Physica A 537 (2020) 122686.

[25] K.Nagel, M.Schreckenberg, A cellular automaton model for freeway traffic, Journal de Physique I 2(1992) 2221-2229.

[26] A.Schadschneider and M.Schreckenberg, Traffic flow models with 'slow-to-start' rules, Ann.Phys. 6(1997) 541.

[27] B.S.Kerner, H.Rehborn, Experimental features and characteristics of traffic jams, Physical Review E 53(1996) R1297-R1300.

[28] F.L.Hall and K.Agyemang-Duah, Freeway capacity drop and the definition of capacity, Transp.Res.Rec 1320 (1991) 91-98.

[29] R.Barlovic, L.Santen,A.Schadschneider, et al., Metastable States in Cellular Automata for Traffic Flow, The European Physical Journal B 5(1998) 793-800.

[30] Luc Int Panis, Steven Broekx, Ronghui Liu, Modelling instantaneous traffic emission and the influence of traffic speed limits, Sci. Total Environ 371 (2006) 270–285.

[31] R.Barlovic and A.Schadschneider and M. Schreckenberg. Random walk theory of jamming in a cellular automaton model for traffic flow[J]. Physica A: Statistical Mechanics and its Applications, 2001.

[32] K.Nassab, M.Schreckenberg, S.Ouaskit, et al., 1/f noise in a cellular automaton model for traffic flow with open boundaries and additional connection sites[J]. Physica A Statal Mechanics & Its Applications, 2005, 354(none):597-605.

[33] N.Jia, S.Ma, Analytical results of the Nagel-Schreckenberg model with stochastic open boundary conditions[J]. Physical Review E Statal Nonlinear & Soft Matter Physics, 2009, 80(4):041105.